\documentclass[prl,aps,amssymb,twocolumn,showpacs,superscriptaddress]{revtex4}
\usepackage{graphicx}

\begin{document}

\title{Phase Sensitive Shot Noise in an Andreev Interferometer} 

\author{B.~Reulet}
\affiliation{Departments of Applied Physics and Physics, Yale University, New Haven CT 06520-8284, USA} 
\affiliation{Laboratoire de Physique des Solides, associ\'e au CNRS,
b\^atiment 510, Universit\'e Paris-Sud 91405 ORSAY Cedex, France}
\author{A.A.~Kozhevnikov}
\author{D.E.~Prober} 
\affiliation{Departments of Applied Physics and Physics, Yale University, New Haven CT 06520-8284, USA} 
\author{W.~Belzig}
\affiliation{Department of Physics and Astronomy, University of Basel, Klingelbergstr. 82,
4056 Basel, Switzerland}
\author{Yu.V.~Nazarov}
\affiliation{Department of Applied Physics and Delft Institute of
Microelectronics and Submicron technology, 
Delft University of Technology, Lorentzweg 1, 2628 CJ Delft, The Netherlands}

\date{\today} 

\begin{abstract}
We investigate nonequilibrium noise in a diffusive 
Andreev interferometer, in which currents emerging from two Normal metal/Superconductor (N-S) interfaces can interfere. We observe a modulation of the shot noise when the phase difference between the two N-S interfaces is varied by a magnetic flux. This is the signature of phase-sensitive fluctuations in the normal metal. The effective charge inferred from the shot noise measurement is close to $q_{eff}=2e$ but shows phase-dependent deviations from $2e$ at finite energy, which we interprete as due to pair correlations.
Experimental data are in good agreement with theoretical predictions based on an extended Keldysh Green's function approach.
\end{abstract}
\pacs{72.70.+m, 73.23.Ps, 42.50.Lc, 05.40.+j}
\maketitle


Transport in mesoscopic normal metal - superconductor (N-S) structures has attracted great interest recently \cite{NSreview} due to the subtle and varied ways in which coherence is exhibited in these systems. Propagation of the superconducting correlations into the normal metal is realized via Andreev reflection~\cite{Andreev} at the N-S boundary: an electron can cross the N-S boundary from the normal metal only by leaving behind a hole correlated to the electron. This transfers a charge of $2e$. The existence of electron-hole correlated pairs in the normal metal is a consequence of the presence of the superconducting reservoir.  This proximity effect strongly affects all the properties of the normal metal: its thermodynamics (e.g. the existence of a supercurrent in a normal wire between two S reservoirs), its transport properties (reentrant conductance of an Andreev wire - a diffusive normal wire between N and S reservoirs) and its fluctuations (doubled shot noise in the Andreev wire, a consequence of the charge transfer being $2e$)~\cite{NSreview,doublingexp}. 
  
After an Andreev process, the reflected hole carries information about the phase of the superconducting order parameter of the S reservoir at the N-S interface. When two S reservoirs are connected to the same phase-coherent device, a phase gradient develops along the normal metal, resulting in phase-dependent properties. In an Andreev interferometer - a device containing a superconducting loop, all the electronic properties are periodic with the magnetic flux $\Phi$ enclosed by the loop, with a period of the flux quantum, $\Phi_0 = h/(2e)$. This has been observed in the supercurrent and in the conductance\cite{loops,denHartog}, but phase-dependent fluctuations have not been reported.  Studies of such non-equilibrium noise (shot noise) are of interest to elucidate the correlations of the charge transfer process. 

In this work we present the results of measurements and modeling of conductance and non-equilibrium current noise, $S_I$, in an Andreev interferometer.  $S_I$ is the spectral density of the current fluctuations.  The effective charge is defined as $q_{eff}=(3/2)(dS_I/dI)$.   At finite energy, $q_{eff}$ is found to be close to its 'standard' value, $2e$, but with a phase sensitive component.  This deviation from $2e$ is due to the anticorrelated entry of pairs into the normal metal, related to their spatial overlap. This is, to our knowledge, the first experimental proof that electronic fluctuations (noise) can be phase sensitive.  The experimental results agree very well with our calculations based on the counting statistics approach~\cite{Nazarov99,Belzig:2001}. 

The theoretical calculation of the conductance of Andreev interferometers is based on the mesoscopic proximity effect theory~\cite{SupLatt,eilenberger,Nazarov:1994,Nazarov:1996} using the quasiclassical Usadel equation.  The resistance exhibits a minimum at an energy of order the Thouless energy, $E_C=\hbar D/L^2$ ($D$ is the diffusion constant, and $L$ here is the length of one arm of our interferometer) ~\cite{Nazarov:1996,denHartog}. This gives rise to the reentrant behavior of the resistance with temperature $T$ ($E\sim k_BT$) or bias voltage $V$ ($E=eV$).
 
\begin{figure}
\includegraphics[width= 0.9\columnwidth]{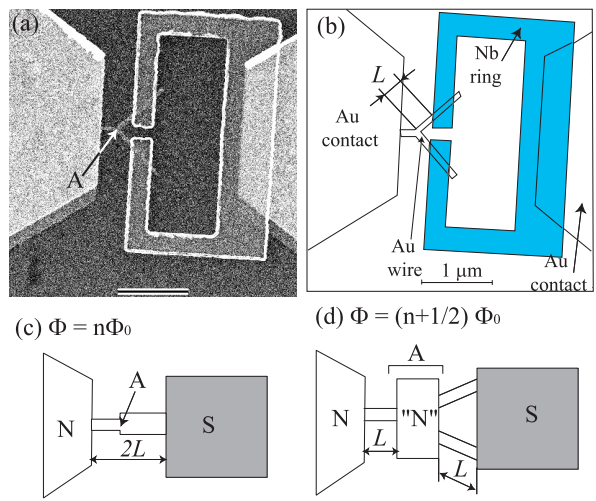}
\caption{(a) SEM picture of the device; (b) the device schematic; (c) and (d) models of the device at integer and half-integer magnetic flux, respectively. N and S are, respectively, normal and superconducting reservoirs.
\vspace{-5mm}
}
\end{figure}

The noise of the Andreev interferometer can be understood by first considering the Andreev wire.  The current spectral density at $T=0$ for both systems is $S_I=(2/3)2eI$ at energies much smaller and much larger than the Thouless energy, for $I>0$~\cite{doubling,doublingexp}.  This is twice the value for a diffusive normal wire between two normal reservoirs at $T=0$.  This doubling of the shot noise can be interpreted as the effective charge being $q_{eff}=2e$ due to Andreev reflection.  The behavior at intermediate energies is more subtle, and was not initially accessible. Recently, methods were developed~\cite{Belzig:2001} to treat the shot noise at arbitrary energies and obtain the full counting statistics. It was shown for the Andreev wire that the noise in the range $E\sim E_C$ has a non-trivial energy dependence~\cite{Belzig:2001,doublingexp}.  However, the physical interpretation of this behavior was not evident.  For interferometers, more complex, phase-dependent behavior is predicted; see below.  The ability to vary the phase has allowed us to test separately various aspects of the theory.  In particular, we have established in the present work that the non-trivial energy dependence of the shot noise at intermediate energies, seen for Andreev wires and interferometers, arises from correlations between the electron-hole pairs entering the normal metal, and that phase gradients suppress this effect.

Devices were fabricated and measured at Yale University. An SEM picture and a schematic of the device studied are presented in Fig.~1a (two similar devices were studied). The device is a diffusive Au wire, shaped like a "Y", in contact with a thick Au reservoir, and in contact with 2 terminals of a large Nb loop. The Nb loop is attached to another Au reservoir. The conductance of the structure is measured between the two Au reservoirs. The phase difference $\varphi$ between the superconducting terminals is controlled by application of a magnetic field perpendicular to the plane of the superconducting loop; $\varphi=2\pi\Phi/\Phi_0$. The thin Au wire and the thick Au reservoirs were deposited using a double-angle evaporation technique~\cite{FultonDolan} in a single vacuum pump down. The surface of the Au was then ion-beam cleaned before Nb deposition to ensure a transparent N-S interface. The Au wire is about 10 nm thick and has an almost temperature-independent sheet resistance of $\sim15~\Omega/\square$, which corresponds to an electron diffusion constant $D\approx3.3\;10^{-3}$~m$^2$/s. The Au reservoir is $70$~nm thick and has a sheet resistance of $\sim0.5~\Omega/\square$. The Nb film is $80$~nm thick. 
For our device, we define the Thouless energy $E_C=\hbar D/L^2 \approx~30\mu$eV, 
where $L\approx~270$nm is the length of each of the 3 sections of the device. 
Measurements were performed in a dilution refrigerator at a mixing chamber 
temperature $T=43$~mK. At low temperature the electron energy relaxation is dominated by electron-electron interactions~\cite{Altshuler:1982} and the associated inelastic length $L_{ee}$ is larger than $L$, so the transport in the device is elastic~\cite{Lee}.

The differential resistance, $R_{diff}=dV/dI$ was measured at $\sim200$Hz as a function of bias
voltage $V$ for several values of magnetic flux, using lines carefully filtered with cryogenic low-pass filters.
The current fluctuations $S_I$ in the sample were measured in a frequency band $\Delta f$ from $1.25$ to $1.75$GHz using a cryogenic HEMT amplifier. The noise emitted by the sample passes through a cold circulator, to isolate the sample from amplifier emissions, and is then amplified by the cryogenic amplifier and rectified at room temperature after further amplification. The detected power is thus given by $P_{det}=G\Delta f(k_BT_{out}+k_BT_A)$ where $G$ is the gain of the amplifier chain, $T_A\sim6.5$K is the noise temperature of the amplifier, and $T_{out}$ is the effective temperature corresponding to the noise power coming from the sample ($T_{out}=0.04-0.6$K for $V=0-150\;\mu$V). We determine $G\Delta f$ and $T_A$ by measuring the sample's Johnson noise vs.~temperature at $V=0$ and its shot noise at $eV\gg(k_BT,E_C)$.
We modulate the current through the sample to suppress the contribution of $T_A$, and measure $dP_{det}/dI$. This gives $dT_{out}/dI$. $T_{out}$ is given by $T_{out}=(1-\Gamma^2)T_N+\Gamma^2T_{in}$, and $S_I$ is given by $S_I=4k_BT_N/R_{diff}$. Here $T_N$ is the sample's noise temperature and $R_{diff}$ the differential resistance at the measurement frequency~\cite{noteRdiff}, $\Gamma^2$ is the power reflection coefficient of the sample and $T_{in}$ the external noise incoming to the sample. In the formula for $T_{out}$, the first term on the right represents the noise emitted by the sample which is coupled to the amplifier. The second term represents the external noise the sample reflects. It has a magnitude of a few percent of the first term~\cite{gamma2}.
In the data analysis we take $\Gamma^2=0$, since measurements of the relative variation of $\Gamma^2(\Phi,V)$ show  that taking $\Gamma^2=0$ leads to an error of $<0.01e$ for the effective charge.

\begin{figure}
\includegraphics[width=\columnwidth]{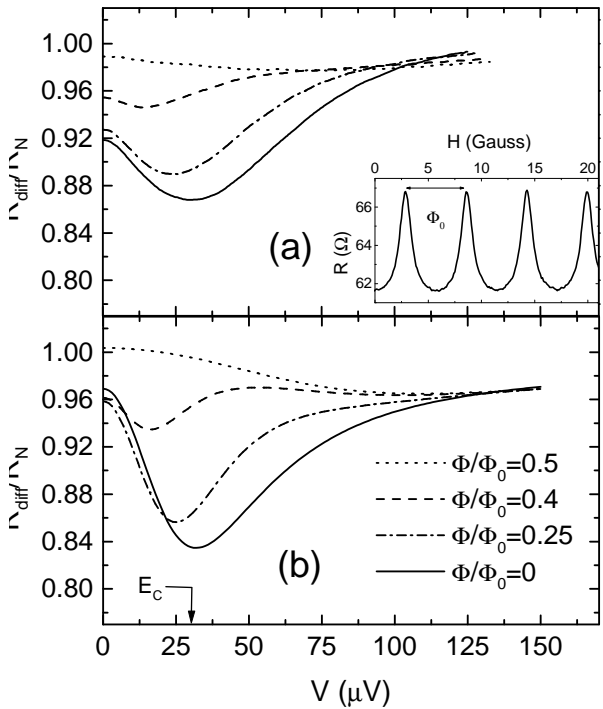}
\vspace{-1.5cm}
\caption{(a) Experimental data of the differential resistance vs. bias voltage at different values of magnetic flux (mixing chamber temperature $T=43$mK). The inset shows magnetic field dependence of the zero-bias differential resistance. (b) Theoretical predictions for $E_C=30\mu$eV and an electron temperature of $T=43$mK.
\vspace{-5mm}
}
\end{figure}

Measurements of $R_{diff}$ vs. bias voltage are given in Fig. 2a. $R_{diff}$ has a systematic dependence on the magnetic flux. At integer flux, $\Phi=n\Phi_0$, the reentrant behavior of the differential resistance vs.~bias voltage is
pronounced.  The maximum dip in resistance below the value at large voltage, $R_N$, is about $13$\%. $R_{diff}(V=0)$ is lower than its normal state value ($R_N=67.5\Omega$ taken at $V=400\;\mu$eV) possibly because of finite temperature and finite phase coherence. The reentrant behavior is also present at half-integer 
flux, but the amplitude of the resistance change is smaller, about $2$\%. 
The minimum of $R_{diff}$ occurs at a voltage $V\approx E_{C}/e$ for
$\Phi=n\Phi_0$ and at a substantially higher bias voltage, $V\approx3E_{C}/e$, for half-integer flux. 
The magnetic field dependence of the differential resistance at $V=0$ is shown in the inset of Fig~2a. The observed magnetic field period is within $10$\% equal to $\Phi_0/A$, with $A$ the inner area of the loop.

We calculated $R_{diff}$ using the approach
of Ref.~\onlinecite{Belzig:2001}; see Fig.~2b. 
The value of the Thouless energy used in the calculations, $E_C=30\;\mu$eV, is inferred from the 
sample geometry. The electron temperature is assumed to be $43$~mK ($k_BT=4\mu eV$), that of the mixing chamber. No fitting parameters were used. Use of a slightly higher temperature would somewhat improve the fit but not change the qualitative behavior. Our calculation does not include the effects of inelastic scattering, imperfect interfaces nor additional heating of the reservoirs at finite voltages. Even so, the conductance of the Andreev interferometer is seen to be well described by theory, in its basic form. This gives us confidence  that the sample parameters and geometry are correctly described.

The behavior of $R_{diff}$ seen in Fig. 2 can be understood qualitatively.  The dip of $R_{diff}$ for $\Phi=0$ at $E\approx E_C$ is like that seen for an Andreev wire of length $2L$. (For our Y-shaped interferometer, the two right hand arms of length $L$ can be taken to be in parallel, see Fig. 1c.)  At high energy $R_{diff}=R_N$.  The dip is due to a competition between the effect of induced, correlated electron-hole pairs at finite energy, which reduce the resistance, and the formation of a gap at yet lower energies, which increases the resistance.  For $\Phi=\Phi_0/2$, point A of the interferometer (Fig. 1d) is driven normal so that the length affected by the superconductor is $L$.  Thus, the dip occurs at a voltage roughly four times larger.  Also, the resistance of those two arms is a fraction, $1/3$, of $R_N$, so the dip is much smaller than for $\Phi=0$.

\begin{figure}
\includegraphics[width=\columnwidth]{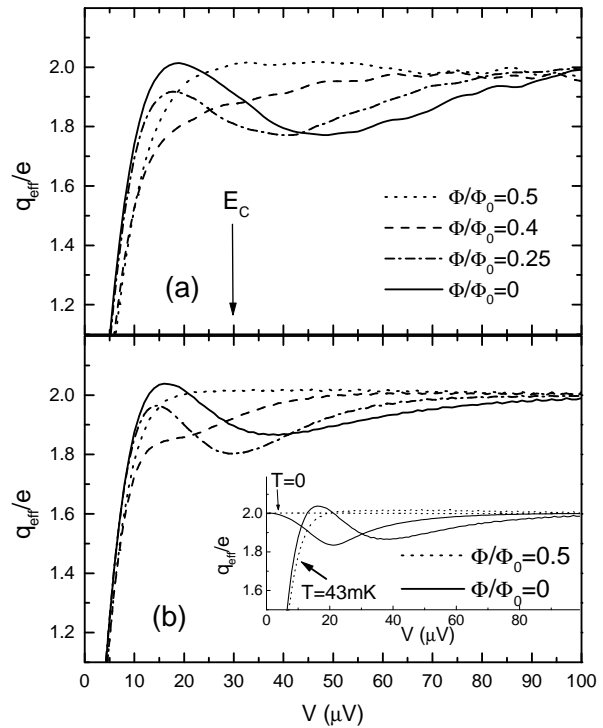}
\vspace{-1cm}
\caption{(a) Experimentally measured effective charge $q_{eff}$ for several values of magnetic flux. (b) Theoretical predictions for $E_C=30\mu$eV and $T=43$mK. The inset shows the theory for $\Phi=0$ and $\Phi=\Phi_0/2$ at $43$mK and for $T=0$.
\vspace{-5mm}
}
\end{figure}

From the noise measurements we deduce the effective charge, $q_{eff}=(3/2)(dS_I/dI)$; see Fig. 3a. At finite energy ($E>k_BT$) the effective charge reflects the charge transferred but also includes the effects of correlations in the transfer process.
By considering $dS_I/dI$ rather than $dS_I/dV$ we eliminate the trivial effect of a non-linear $I(V)$ characteristic. The voltage dependence of $dS_I/dI$ yields information on energy-dependent correlations between charge transfers.
Fig. 3b gives the theory results based on full counting statistics. The inset shows the theory for $\Phi=0$ and $\Phi=\Phi_0/2$, for $T=43$mK and $T=0$.
The effective charge is seen in the theory to be independent of the phase difference
at bias voltages larger than $\sim100~\mu$V, with significant phase modulation of 
$q_{eff}$  in the bias voltage range $10-80~\mu$V. The maximum magnitude
of the observed dip of $q_{eff}$ vs. voltage is $\sim10$\%, and occurs for $\Phi\sim\Phi_0/4$. There is no dip for $\Phi=\Phi_0/2$. For $T=0$, $q_{eff}$ returns to $2e$ as $V\rightarrow0$. At finite temperature, $q_{eff}$ goes to zero for $eV\ll k_BT$. This is because $S_I$ reduces to Johnson noise at $V=0$. Thus, the decrease of $q_{eff}$ at very low voltages is not related to Andreev physics. In contrast, the dip near $E_C$ is due to the energy dependence of Andreev reflection.

The experimental results are in fairly good agreement with the theoretical predictions. As expected, there is no phase modulation of $q_{eff}$ at large energies $eV\gg E_C$, and here $q_{eff}=2e$. At $E\sim E_C$, the effective charge is smaller for integer flux than for half-integer
flux. The non-trivial energy/flux dependence predicted (crossings of the different curves) is seen in the experiment, though the agreement is not perfect. The magnitude of the dip of $q_{eff}$ in the data is also close to the theoretical prediction.

To understand the origin of the dip of $S_I$ seen for $\Phi=0$, we have also solved a generalized Boltzmann-Langevin (BL) equation. In such an approach correlations due to the superconductor enter  through the energy- and space-dependent conductivity, which gives $I(V,\Phi)$. At $T=0$, the BL result for all flux values is simply $S_I^{BL}=(2/3)2eI(V,\Phi)$, i.e., $q_{eff}=2e$ at all energies. This implies that the deviation of the effective charge from $2e$ must be due to fluctuation processes which are not related to single-particle scattering, on which the BL approach is based.  
We believe that the higher-order process which is responsible for the dip of $S_I$ is a two-pair correlation process.  At high energies ($E>E_C$) the electron-hole pair states have a length $\sim(\hbar D/E)^{1/2}$, shorter than $2L$. This results in uncorrelated entry of pairs into the normal region.  For $E<E_C$ the pair size is larger, and the spatial overlap prevents fully random entry, suppressing $S_I$. Suppressed shot noise is a signature of anti-correlated charge entry~\cite{BuBlan}. At lower energies (at $T=0$) the effective charge is predicted to return to 2e; we do not yet have a physical interpretation of this. In any case, for the case of $\Phi=\Phi_0/2$, the dip of $q_{eff}$ is fully suppressed, according to the theory. This means that the phase gradients destroy the pair correlation effect.  

In summary we have understood the energy- and phase dependence of the shot noise of an Andreev interferometer. The correlation of pair entry into the normal metal causes deviations from the effective charge of $q_{eff}=2e$.
The effects of pair correlations and of phase gradients on other properties will be explored in future experiments.


The authors wish to acknowledge I.~Siddiqi, C.~Wilson and L.~Frunzio for
assistance with device fabrication and A. Clerck, M.~Devoret and R.~Schoelkopf for useful discussions. This work was supported by NSF DMR grant 0072022. The work of W.B. was supported by the Swiss NSF and the NCCR Nanoscience.


\begin{thebibliography}{99}
\bibitem{Lee} We observe significant harmonic content of the $R$ vs.~flux curve (inset of Fig. 2a) This implies a phase-coherence length $>800$nm, i.e. much larger than the sample. Previous measurements on a longer interferometer ($L\sim1\mu$m) also gave evidence that $L_{ee}>L$ in that device.
\bibitem{noteRdiff} In the data analysis we have replaced $R_{diff}$ by $dV/dI$ of Fig. 2(a). In our short phase-coherent sample transport is elastic. Thus the ac conductance is given by the DC $I(V)$ characteristics shifted by $\pm\hbar\omega/e\approx\pm6\mu$V~\cite{tunnel}. Since the characteristic scale for $dV/dI$ is $E_C$ at $\Phi=0$, finite frequency corrections to  $R_{diff}$ should be small. In the same way, finite frequency corrections to the shot noise involve $S_I(V\pm\hbar\omega/(2e))$
~\cite{BuBlan}. Since $S_I(V)$ has no structure on a scale $\hbar\omega/2e\approx3\mu$V, the only effect of the finite frequency is to slightly broaden the low voltage rise of $dS_I/dI$ leading to an apparent temperature slightly higher than that of the electrons.
\bibitem{gamma2} The sample is reasonably well matched to the rf system ($\Gamma^2=(R_{diff}-50\Omega)^2/(R_{diff}+50\Omega)^2\sim2\%$) and we employ an circulator at $T=50$mK which attenuates by 20dB the noise emitted by the amplifier towards the sample ($T_{emit}\sim2K$). At $V=0$, $R_{diff}$ and $\Gamma^2$ are flux dependent, and $T_N=T$; yet $T_{out}$ at $V=0$ does not depend on the flux, confirming that it is not affected by $T_{in}$.
\end{thebibliography}
\end{document}